\documentclass[preprint,prl,aps,showpacs,showkeys,superscriptaddress]{revtex4}

\usepackage[latin1]{inputenc}
\usepackage{amssymb, amsmath}
\usepackage[dvips,final]{graphicx}
\usepackage[english]{babel}
\usepackage{bm}
\usepackage{subfigure}
\usepackage{color}
\usepackage{amssymb}
\usepackage{amsmath}
\usepackage{graphicx}

\begin{document}

\title{Influences of excitation-dependent bandstructure changes on InGaN
light-emitting diode efficiency }
\title{Influences of excitation-dependent bandstructure changes on InGaN
light-emitting diode efficiency }
\author{Weng W. Chow}
\email{wwchow@sandia.gov}

\begin{abstract}
Bandstructure properties in wurtzite quantum wells can change appreciably
with changing carrier density because of screening of quantum-confined Stark
effect. An approach for incorporating these changes in an InGaN
light-emitting-diode model is described. Bandstructure is computed for
different carrier densities by solving Poisson and k$\cdot $p equations in
the envelop approximation. The information is used as input in a dynamical
model for populations in momentum-resolved electron and hole states.
Application of the approach is illustrated by modeling device internal
quantum efficiency as a function of excitation.
\end{abstract}
\date{\today}
\pacs{todo}
\maketitle
\address{ Sandia National Laboratories, Albuquerque, New Mexico 87185-1086,
USA }
\section{Introduction}

Considerable progress is being made in advancing InGaN light-emitting diodes
(LEDs). However, there are still concerns involving performance limitations.
An example is efficiency loss at high current density (efficiency droop) 
\cite{krames}, which can limit use of LEDs in applications requiring intense
illumination. Understanding and mitigating the efficiency droop mechanism is
important. Several explanations have been proposed, including carrier
leakage \cite{schubert}, Auger recombination \cite{shen}, junction heating 
\cite{efremov}, carrier and defect delocalizations \cite{chichibu,hader1}.
The assertions are much debated. For example, in the case of Auger
scattering, discrepancy exists in the Auger coefficient estimation between
experimental-curve fitting and microscopic calculations \cite%
{shen,ryu,hader2,dellaney}.

Discussions involving InGaN LED efficiency are commonly based on a rate
equation for the total carrier density. The approach allows one to describe
radiative and nonradiative carrier loss rates, where the latter typically
includes ad-hoc terms for producing an efficiency droop. A particularly
successful model, in terms of reproducing experimental efficiency versus
injection current data, is the $ABC$ model. \cite{shen,ryu} The model's name
derives from the three phenomenological constants ($A$, $B$ and $C$)
introduced to account for Shockley-Read-Hall (SRH), radiative-recombination
and Auger-scattering carrier losses, respectively.\ Bandstructure effects
enter indirectly via these coefficients.

It is known that the bandstructure in wurtzite quantum-well (QW) structures
can change noticeably with carrier density because of screening of the
quantum-confined Stark effect (QCSE) \cite{shur,hangleiter}.\ Incorporating
these changes into the $ABC$ model is challenging, without compromising the
attractiveness of having only three fitting parameters, each with direct
correspondence to a physical mechanism. This paper considers an alternative
that allows direct input of bandstructure properties, in particular, the
band energy dispersions, confinement energies and optical transition matrix
elements, as well as their carrier-density dependences arising from
screening of piezoelectric and spontaneous polarization fields. The model
has the further advantage of providing a consistent treatment of spontaneous
emission, carrier capture and leakage, and nonequilibrium effects. Thus, the
fitting parameter, $B$ is eliminated and effects, such as plasma heating,
are taken in account within an effective relaxation rate approximation for
carrier-carrier and carrier-phonon scattering. All this is accomplished by
extending a previously reported non-equilibrium LED\ model that is based on
dynamical equations for electron and hole occupations in each momentum ($k$)
state \cite{chow1}. The additions include an algorithm for simplifying and
extracting bandstructure information relevant to the dynamical equations.
Detailed bandstructure properties are obtained from solving $k\cdot p$ and
Poisson equations \cite{chuang}. Furthermore, since distinction between QW
and barrier states is sometimes difficult in the presence of strong internal
electric fields, extension is made to treat optical emission from these
states on equal footing.

Section 2 describes the model, derivation of the working equations and
calculation of input bandstructure properties.\ Section 3 demonstrates the
application of the $k-$resolved model by calculating internal quantum
efficiency (IQE) as a function of injection current for a multi-QW InGaN
LED. Results are presented to illustrate IQE behavior that may be overlooked
when not accounting for the excitation dependences of bandstructure. Most
interesting is a possible contribution to efficiency droop from a change in
relative emission contributions from QWs and barriers.\ Simulation results
are presented to demonstrate the robustness of this mechanism to input
parameter variations.\ Also in this section, a back-of-the-envelop
derivation is used to associate the predicted droop to bandstructure changes
from screening of QCSE. Section 4 further explains the bandstructure-induced
droop mechanism by discussing the changes in QW confinement energies and
envelop function overlap with increasing excitation. The section also
presents a possible resolution to the discrepancy on the Auger coefficient
estimation between $ABC$ model and microscopic calculation. With the present
model, the Auger coefficient necessary to maintain an efficiency droop at
high injection current, as presently observed in experiments, is in the
range of $5\times 10^{-32}$ to $10^{-31}cm^{6}s^{-1}$, which is in closer
agreement with microscopic calculations.\ Section 5 summarizes the paper.

\section{Theory}

The following Hamiltonian, adapted from quantum optics \cite{jaynes}, is
used in the derivation of spontaneous emission from QW and barrier
transitions:%
\begin{equation}
H=\sum_{i}\varepsilon _{i}^{e}a_{i}^{\dagger }a_{i}+\sum_{j}\varepsilon
_{j}^{h}b_{j}^{\dagger }b_{j}+\sum_{q}\hbar \Omega _{q}c_{q}^{\dagger
}c_{q}-\sum_{i,j,q}\wp _{ij}\sqrt{\frac{\hbar \Omega _{q}}{V\epsilon _{b}}}%
(a_{i}b_{j}c_{q}^{\dagger }+c_{q}b_{j}^{\dag }a_{i}^{\dag })  \label{1}
\end{equation}%
The summations are over QW and barrier states with subscript $i(j)$
representing $e,\alpha _{e},k_{\perp }$ ($h,\alpha _{h},k_{\perp }$) for QW
states and $e,k$ ($h,k$) for barrier states. In this notation, each QW state
is denoted by its charge $\sigma $, subband $\alpha _{\sigma }$ and in-plane
momentum $k_{\perp }$. A bulk state is specified by its charge $\sigma $ and
3-dimensional carrier momentum $k$. In Eq. (\ref{1}), $a_{i}$,$%
a_{i}^{\dagger }$ ($b_{j}$,$b_{j}^{\dagger }$) are electron (hole)
annihilation and creation operators, $c_{q}$, $c_{q}^{\dagger }$\ are
corresponding operators for the photons, $\varepsilon _{i}^{\sigma }$ is the
carrier energy, $\Omega _{q}$ is the photon frequency, $\wp _{ij}$ is the
dipole matrix element, $V$ is the active region volume and $\epsilon _{b}$
is the host permittivity. Using the Hiesenberg operator equations of motion
and the above Hamiltonian, the carrier populations and polarizations evolve
according to%
\begin{eqnarray}
\frac{\text{d}\left\langle a_{i}^{\dagger }a_{i}\right\rangle }{\text{d}t}
&=&\cos i\sum_{j,q}\wp _{ij}\sqrt{\frac{\Omega _{q}}{\hbar V\epsilon _{b}}}%
\left[ \left\langle c_{q}b_{j}^{\dag }a_{i}^{\dag }\right\rangle
-\left\langle a_{i}b_{j}c_{q}^{\dagger }\right\rangle \right]  \label{2} \\
\frac{\text{d}\left\langle b_{j}^{\dagger }b_{j}\right\rangle }{\text{d}t}
&=&i\sum_{i,q}\wp _{ij}\sqrt{\frac{\Omega _{q}}{\hbar V\epsilon _{b}}}\left[
\left\langle c_{q}b_{j}^{\dag }a_{i}^{\dag }\right\rangle -\left\langle
a_{i}b_{j}c_{q}^{\dagger }\right\rangle \right]  \label{3}
\end{eqnarray}%
\begin{eqnarray}
\frac{\text{d}\left\langle a_{i}b_{j}c_{q}^{\dagger }\right\rangle }{\text{d}%
t} &=&-i\left( \Omega _{q}-\Omega _{i}{}_{j}\right) \left\langle
a_{i}b_{j}c_{q}^{\dagger }\right\rangle  \notag \\
&&+i\wp _{ij}\sqrt{\frac{\Omega _{q}}{\hbar V\epsilon _{b}}}\left\langle
\left( a_{i}^{\dagger }a_{i}+b_{j}^{\dagger }b_{j}-1\right) c_{q}^{\dagger
}c_{q}+a_{i}^{\dagger }a_{i}b_{j}^{\dagger }b_{j}\right\rangle +...
\label{4}
\end{eqnarray}%
where $\Omega _{ij}=\left( \varepsilon _{i}^{e}+\varepsilon _{j}^{h}\right)
/\hbar $ is the transition frequency. Factorizing the operator products and
truncating at the first level (Hartree-Fock approximation) give for Eq. (\ref%
{4})

\begin{align}
\frac{\text{d}\left\langle a_{i}b_{j}c_{q}^{\dagger }\right\rangle }{\text{d}%
t}& =i\left( \Omega _{q}-\Omega _{i}{}_{j}\right) \left\langle
a_{i}b_{j}c_{q}^{\dagger }\right\rangle  \notag \\
& -i\wp _{ij}\sqrt{\frac{\Omega _{q}}{\hbar V\epsilon _{b}}}\left[ \left(
\left\langle a_{i}^{\dagger }a_{i}\right\rangle +\left\langle b_{j}^{\dagger
}b_{j}\right\rangle -1\right) \left\langle c_{q}^{\dagger
}c_{q}\right\rangle +\left\langle a_{i}^{\dagger }a_{i}\right\rangle
\left\langle b_{j}^{\dagger }b_{j}\right\rangle \right]  \label{5}
\end{align}%
For an LED, it is customary to assumed that cavity influence is sufficiently
weak so that $\left\langle c_{q}^{\dagger }c_{q}\right\rangle <<1$ and only
the spontaneous emission contribution is kept. Additionally, polarization
dephasing is introduced, where the dephasing (with coefficient $\gamma )$ is
assumed to be considerably faster than the population changes. This allows
integration of Eq. (\ref{5}). The result is used to eliminate the
polarization in Eqs. (\ref{2}) and (\ref{3}), giving

\begin{eqnarray}
\frac{\text{d}\left\langle a_{i}^{\dagger }a_{i}\right\rangle }{\text{d}t}
&=&-\left\langle a_{i}^{\dagger }a_{i}\right\rangle \sum\limits_{j,q}\frac{%
2\Omega _{q}}{\hbar \epsilon _{b}V\gamma }\left\vert \wp _{ij}\right\vert
^{2}\left\langle b_{j}^{\dagger }b_{j}\right\rangle \left[ 1+\left( \frac{%
\Omega _{ij}-\Omega _{q}}{\gamma }\right) ^{2}\right] ^{-1}  \label{6} \\
\frac{\text{d}\left\langle b_{j}^{\dagger }b_{j}\right\rangle }{\text{d}t}
&=&-\left\langle b_{j}^{\dagger }b_{j}\right\rangle \sum\limits_{i,q}\frac{%
2\Omega _{q}}{\hbar \epsilon _{b}V\gamma }\left\vert \wp _{ij}\right\vert
^{2}\left\langle a_{i}^{\dagger }a_{i}\right\rangle \left[ 1+\left( \frac{%
\Omega _{ij}-\Omega _{q}}{\gamma }\right) ^{2}\right] ^{-1}  \label{7}
\end{eqnarray}%
Converting the photon momentum summation into an integral, i.e.

\begin{equation}
\sum\limits_{q}\rightarrow 2\frac{V}{\left( 2\pi \right) ^{3}}%
\int\limits_{0}^{\infty }\text{d}q\ 4\pi q^{2}  \label{7a}
\end{equation}%
where $\Omega _{q}=qc$ and $c$ is the speed of light in the semiconductor,
the right-hand sides of Eqs. (\ref{6}) and (\ref{7}) may be integrated to
give%
\begin{eqnarray}
\frac{\text{d}\left\langle a_{i}^{\dagger }a_{i}\right\rangle }{\text{d}t}
&=&-\left\langle a_{i}^{\dagger }a_{i}\right\rangle \sum\limits_{j}\frac{%
n_{b}}{\hbar \epsilon _{0}\pi c^{3}}\left\vert \wp _{ij}\right\vert
^{2}\Omega _{ij}^{3}\left\langle b_{j}^{\dagger }b_{j}\right\rangle
\label{8} \\
\ \frac{\text{d}\left\langle b_{j}^{\dagger }b_{j}\right\rangle }{\text{d}t}
&=&-\left\langle b_{j}^{\dagger }b_{j}\right\rangle \sum\limits_{i}\frac{%
n_{b}}{\hbar \epsilon _{0}\pi c^{3}}\left\vert \wp _{ij}\right\vert
^{2}\Omega _{ij}^{3}\left\langle a_{i}^{\dagger }a_{i}\right\rangle
\label{9}
\end{eqnarray}

Writing explicitly for QW populations and adding phenomenologically SRH
carrier loss and relaxation contributions from carrier-carrier and
carrier-phonon scattering, gives

\begin{eqnarray}
\frac{\text{d}n_{\sigma ,\alpha _{\sigma },k_{\bot }}}{\text{d}t}
&=&-n_{\sigma ,n_{\sigma },k_{\bot }}\sum\limits_{\alpha _{\sigma ^{\prime
}}}b_{\alpha _{\sigma },\alpha _{\sigma ^{\prime }},k_{\bot }}n_{\sigma
^{\prime },\alpha _{\sigma ^{\prime }},k_{\perp }}-An_{\sigma ,n_{\sigma
},k_{\bot }}  \notag \\
&&-\gamma _{c-c}\left[ n_{\sigma ,n_{\sigma },k_{\bot }}-f\left( \varepsilon
_{\sigma ,k_{\perp }},\mu _{\sigma },T\right) \right]  \notag \\
&&-\gamma _{c-p}\left[ n_{\sigma ,n_{\sigma },k_{\bot }}-f\left( \varepsilon
_{\sigma ,k_{\perp }},\mu _{\sigma }^{L},T_{L}\right) \right]  \label{10}
\end{eqnarray}%
where $\sigma ,\sigma ^{\prime }$ is $e,h$ or $h,e$. In Eq. (\ref{10}), $%
\gamma _{c-c}$ and $\gamma _{c-p}$ are the effective carrier-carrier and
carrier-phonon collision rates, respectively, and%
\begin{equation}
b_{\alpha _{\sigma },\alpha _{\sigma ^{\prime }},k_{\bot }}=\frac{1}{\hbar
\epsilon _{b}\pi c^{3}}\left\vert \wp _{\alpha _{\sigma },\alpha _{\sigma
^{\prime }},k_{\bot }}\right\vert ^{2}\Omega _{\alpha _{\sigma },\alpha
_{\sigma ^{\prime }},k_{\bot }}^{3}  \label{10a}
\end{equation}%
where $\wp _{\alpha _{\sigma },\alpha _{\sigma ^{\prime }},k_{\bot }}$ and $%
\Omega _{\alpha _{\sigma },\alpha _{\sigma ^{\prime }},k_{\bot }}$ are the
QW dipole matrix element and transition energy.\ Similarly, for the barrier
populations,%
\begin{eqnarray}
\frac{\text{d}n_{\sigma ,k}^{b}}{\text{d}t} &=&-b_{k}n_{e,k}^{b}n_{h,k}^{b}+%
\frac{J}{eN_{\sigma }^{p}}f\left( \varepsilon _{\sigma ,k}^{b},\mu _{\sigma
}^{p},T_{p}\right) \left( 1-n_{\sigma ,k}^{b}\right) -\gamma _{b}n_{\sigma
,k}  \notag \\
&&-\gamma _{c-c}\left[ n_{\sigma ,k}^{b}-f\left( \varepsilon _{\sigma
,k}^{b},\mu _{\sigma },T\right) \right]  \notag \\
&&-\gamma _{c-p}\left[ n_{\sigma ,k}^{b}-f\left( \varepsilon _{\sigma
,k}^{b},\mu _{\sigma }^{L},T_{L}\right) \right]  \label{11}
\end{eqnarray}%
where

\begin{equation}
b_{k}=\frac{1}{\hbar \epsilon _{b}\pi c^{3}}\left\vert \wp _{k}\right\vert
^{2}\Omega _{k}^{3},  \label{11a}
\end{equation}%
$\wp _{k}$ and $\Omega _{k}$ are the barrier dipole matrix element and
transition energy. In Eq. (\ref{11}) is a pump contribution, where $J$ is
the current density, $e$ is the electron charge, $N_{\sigma
}^{p}=\sum_{k}f\left( \varepsilon _{\sigma ,k}^{b},\mu _{\sigma
}^{p},T_{p}\right) $ and $f\left( \varepsilon _{\sigma ,k}^{b},\mu _{\sigma
}^{p},T_{p}\right) $, the injected carrier distribution, is a Fermi-Dirac
function with chemical potential $\mu _{\sigma }^{p}$ and temperature $T_{p}$%
. For the asymptotic Fermi-Dirac distributions approached via
carrier-carrier collisions, the chemical potential $\mu _{\sigma }$ and
plasma temperature $T$ are determined by conservation of carrier density and
energy. In the case of carrier-phonon collisions, the chemical potential $%
\mu _{\sigma }^{L}$ is determined by conservation of carrier density and the
lattice temperature $T_{L}$ is an input quantity. Total carrier density and
energy are computed by converting the sum over states to integrals, i.e., 
\begin{equation}
\sum\limits_{k_{\perp }}\rightarrow \frac{S}{\left( 2\pi \right) ^{2}}%
2\int\limits_{0}^{\infty }\text{d}k_{\perp }\ 2\pi k_{\perp }\text{ \ and \ }%
\sum\limits_{k}\rightarrow \frac{hS}{\left( 2\pi \right) ^{3}}%
2\int\limits_{0}^{\infty }\text{d}k\ 4\pi k^{2}  \label{11b}
\end{equation}%
where $S$ and $h$ are the surface area and thickness of the active active
region consisting of all QW and barrier layers. Further details involving
implementation and comparison with results from quantum-kinetic calculations
are reported elsewhere \cite{chow2,waldmueller}.\ Many-body effects \cite%
{chow3,book} are neglected in Eqs (\ref{10}) and (\ref{11}). Their
incorporation, at least at the level of the screened Hartree-Fock
approximation \cite{chow2}, will be considered in future investigations.

Bandstructure information enters directly into Eqs (\ref{10}) and (\ref{11})
via the dipole matrix elements $\wp _{\alpha _{\sigma },\alpha _{\sigma
^{\prime }},k_{\bot }}$, $\wp _{k}$ and carrier energies $\varepsilon
_{\sigma ,k_{\perp }}$, $\varepsilon _{\sigma ,k}^{b}$. From $k\cdot p$
theory, the QW electron and hole eigenfunctions are \cite{book}%
\begin{equation}
\bigskip \langle r\left\vert \phi _{\sigma ,\alpha _{\sigma },k_{\bot
}}\right\rangle =e^{ik_{\bot }\cdot r_{\bot }}\sum_{m_{\sigma }}\sum_{\beta
_{\sigma }}A_{\beta _{\sigma },\alpha _{\sigma },k_{\bot }}u_{m_{\sigma
},\beta _{\sigma }}(z)\langle r\left\vert m_{\sigma }\right\rangle
\label{12}
\end{equation}%
where $\left\vert m_{\sigma }\right\rangle $ is a bulk electron or hole
state, $u_{m_{\sigma },\beta _{\sigma }}(z)$ is the $\beta _{\sigma }$-th
envelop function associated with the $m_{\sigma }$ bulk state, $A_{\beta
_{\sigma },\alpha _{\sigma },k_{\bot }}$ is the amplitude of the $\beta
_{\sigma }$-th envelop function contributing to the $\alpha _{\sigma }$-th
subband at momentum $k_{\bot }$, z is position in the growth direction and $%
r_{\bot }$ is position in the QW plane \ Using Eq. (\ref{12}), the square of
the dipole matrix element may then be written as

\begin{eqnarray}
\left\vert \wp _{\alpha _{e},\alpha _{h},k_{\perp }}\right\vert ^{2} &\equiv
&\left\vert \left\langle \phi _{e,\alpha _{e},k_{\bot }}\right\vert
ex\left\vert \phi _{h,\alpha _{h},k_{\bot }}\right\rangle \right\vert ^{2} 
\notag \\
&=&\left\vert \wp _{bulk}\right\vert ^{2}\xi _{\alpha _{e},\alpha
_{h},k_{\perp }}  \label{13}
\end{eqnarray}%
where%
\begin{eqnarray}
\xi _{\alpha _{e},\alpha _{h},k_{\perp }} &=&\frac{1}{4}\left\vert
\sum_{\beta _{e}}\sum_{\beta _{h}}\sum_{m_{e}}\sum_{m_{h}}A_{\beta
_{e},\alpha _{e},k_{\bot }}A_{\beta _{h},\alpha _{h},k_{\bot }}\right. 
\notag \\
&&\left. \times \int_{-\infty }^{\infty }dz\ u_{m_{e},\beta
_{e}}(z)u_{m_{h},\beta _{e}}(z)\right\vert ^{2}  \label{13a}
\end{eqnarray}%
and the bulk dipole matrix element in the absence of an electric field is
given by%
\begin{equation}
\left\vert \wp _{bulk}\right\vert ^{2}=\frac{\hbar ^{2}}{2m_{0}\varepsilon
_{g}}\left( \frac{m_{0}}{m_{e}}-1\right) \left( 1+\frac{\Delta _{1}+\Delta
_{2}}{\varepsilon _{g}}\right) \,,  \label{13b}
\end{equation}%
$\varepsilon _{g}$ is the bulk material bandgap energy, $m_{0}$ and $m_{e}$
are the bare and effective electron masses, $\Delta _{1}$ and $\Delta _{2}$
are energy splittings associated with the bulk hole states. An iterative
solution of the $k\cdot p$ and Poisson equations \cite{chuang} is used to
obtain the energies $\varepsilon _{\sigma ,\alpha _{\sigma },k_{\perp }}$
and $\varepsilon _{\sigma ,k}^{b}$ and the overlap integral $\xi _{\alpha
_{e},\alpha _{h},k_{\perp }}$. For these calculations we use the the bulk
wurtzite material parameters listed in Refs. \cite%
{jenkins,wright,wei,ambacher}.

\section{Results}

With the present model, it is necessary to solve the bandstructure and
population problems self consistently. Simultaneous solution of both
problems is very challenging and perhaps unnecessary. The approach used in
this paper is to first take care of the bandstructure part by iteratively
solving the $k\cdot p$ and Poisson equations for a range of carrier
densities. Bandstructure information needed for the population part are $%
\varepsilon _{\sigma ,\alpha _{\sigma },k_{\perp }}$, $\varepsilon _{\sigma
,k}^{b}$ and $\xi _{\alpha _{e},\alpha _{h},k_{\perp }}$ versus total QW
carrier density, $n_{\sigma }^{qw}=S^{-1}\sum_{\alpha _{\sigma },,k_{\perp
}}n_{\sigma ,\alpha _{\sigma },k_{\perp }}$, where the $n_{\sigma }^{qw}$
dependences are from screening of the QW electric field.

To facilitate the solution of the dynamical population equations, the
carrier states are grouped into two categories: those belonging to the QWs
and those belonging to the barriers. The QW states are treated using Eq. (%
\ref{10}) and the barrier states are treated collectively with Eq. (\ref{11}%
). With a high internal electric field, the distinction between QW and
barrier states may be ambiguous. In this paper, the choice is made by
calculating $\int_{QW}$dz$\ \left\vert u_{m_{\sigma },\beta _{\sigma
}}(z)\right\vert ^{2}$, where integral is performed over the QWs. The states
where the integral is greater than a half are grouped as QW states and the
rest as barrier states. For the problem being addressed, which is the
excitation dependence of IQE, the distinction is only important because only
QW transitions are affected by QCSE. For the barrier transitions, the dipole
matrix element in the presence of an internal electric field is approximated
by an average, where each transition is weighted according to the
occupations of the participating states. When solving the population
equations, grouping the barrier states appreciably reduces numerical demand,
which remains substantial because one is still keeping track of a large
number of $k$-states.

The second step involves numerically solving Eqs. (\ref{10}) and (\ref{11})
with the bandstructure quantities updated at each time step according to the
instantaneous value of $n_{\sigma }^{qw}$. When steady state is reached, IQE
is obtained from dividing the rate of carrier (electron or hole) loss via
spontaneous emission by the rate of carrier injection:

\begin{equation}
IQE=\frac{e}{JS}\left( \sum\limits_{\alpha _{e},\alpha _{h},k_{\perp
}}b_{\alpha _{e},\alpha _{h},k_{\bot }}n_{e,\alpha _{e},k_{\perp
}}n_{h,\alpha _{h},k_{\perp
}}+\sum\limits_{k}b_{k}n_{e,k}^{b}n_{h,k}^{b}\right)  \label{14}
\end{equation}

\begin{figure}[tbp]
\centering
\includegraphics[width=0.5\textwidth]{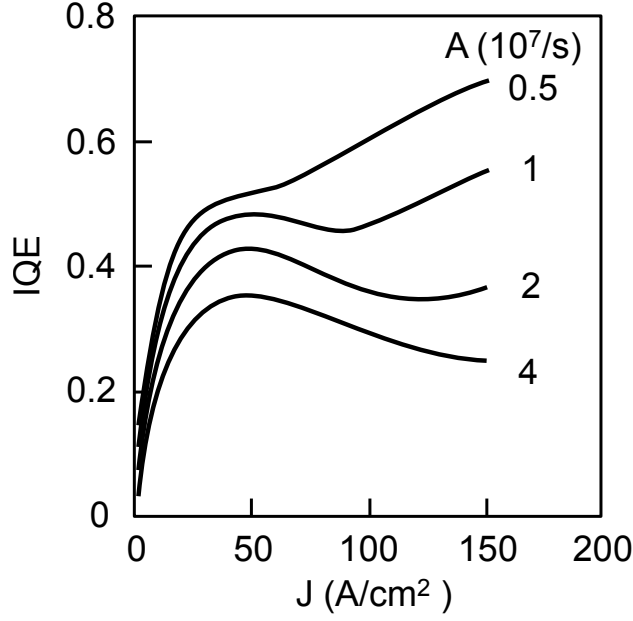}
\caption{Internal quantum efficiency versus current density for different QW
SRH coefficients. The curves are computed using the $k-$resolved model
described by Eqs. (\protect\ref{10}) and (\protect\ref{11}) for a LED with a
In$_{0.2}$Ga$_{0.8}$N/GaN multi-QW active region (see Fig.~\protect\ref%
{fig:4}).}
\label{fig:1}
\end{figure}

Computed IQE\ versus current density curves for different SRH coefficients
in the QWs are plotted in Fig.~\ref{fig:1}. Each curve shows an initial
sharp increase in IQE with injection current, with emission occurring the
instant there is an injected current. Quite interesting, especially because
Auger carrier loss is not included in the model, is the appearance of
efficiency droop in the curves for high $A$ values. A higher SRH coefficient
in QW than barrier is possible in present experimental devices, based on the
roughly three times higher defect density in QWs than in barriers in LEDs
measured at Sandia \cite{andy}. The calculations are performed assuming an
active region consisting five 4nm In$_{0.2}$Ga$_{0.8}$N QWs separated by 6nm
GaN barriers and bounded by 20nm GaN layers. Electric field in the QWs is
determined from the sum of piezoelectric and spontaneous polarization
fields. The electric fields in the barriers are from spontaneous
polarization. Screening of these fields are determined semiclassically
according to Poisson equation, and electron and hole envelop functions.
Input parameters are $A_{b}=10^{7}s^{-1} $, $T_{L}=300K$, $\gamma
_{c-c}=5\times 10^{13}s^{-1}$ and $\gamma _{c-p}=10^{13}s^{-1}$. Effects
arising from doping profile and presence of carrier blocking layers are
ignored.

To uncover the mechanism giving rise to the droop behavior shown in Fig.~\ref%
{fig:1}, it may be more effective to use a less comprehensive model to
isolate bandstructure effects by ignoring carrier leakage and nonequilibrium
effects. Such a model is possible by extending the $ABC$ model to
distinguish between QW and barrier carrier densities, $N_{\sigma }$ and $%
N_{\sigma }^{b}$, respectively. The following phenomenological (and less
rigorous than Eqs .(\ref{10}) and (\ref{11})] rate equations may be written:%
\begin{eqnarray}
\frac{\text{d}N_{\sigma }}{\text{d}t} &=&-BN_{e}N_{h}-AN_{\sigma }
\label{15} \\
\frac{\text{d}N_{\sigma }^{b}}{\text{d}t}
&=&-B_{b}N_{e}^{b}N_{h}^{b}-A_{b}N_{\sigma }^{b}+\frac{J}{eh_{b}}  \label{16}
\end{eqnarray}%
where $\sigma =e$ or $h$. 3-d (volume) densities are used to connect with
the $ABC$ model, especially in terms of the SRH and spontaneous emission
coefficients. Equations (\ref{15}) and (\ref{16}) are coupled by assuming
that intraband collisions are sufficiently rapid so that QW and barrier
populations are in equilibrium at temperature $T$. Defining a total 2-d
carrier density, $N_{2d}=N_{qw}hN_{\sigma }+h_{b}N_{\sigma }^{b}$ allows
combining these equations to give%
\begin{equation}
\frac{\text{d}N_{2d}}{\text{d}t}=-\beta N_{2d}^{2}-A_{b}N_{2d}+\frac{J}{e}
\label{17}
\end{equation}%
where $N_{qw}$ is the number of QWs in the structure, $h$ is the width of
individual QWs,%
\begin{equation}
\beta =\frac{\frac{h}{h_{b}}N_{qw}B+B_{b}\exp \left( \frac{\Delta
_{e}+\Delta _{h}}{k_{B}T}\right) }{\left[ 1+\exp \left( \frac{\Delta _{e}}{%
k_{B}T}\right) \right] \left[ 1+\exp \left( \frac{\Delta _{h}}{k_{B}T}%
\right) \right] }  \label{18}
\end{equation}%
$k_{B}$ is Boltzmann constant and $\Delta _{\sigma }$ is the averaged QW
confinement energy. The steady state solution to Eq. (\ref{17}) gives the
internal quantum efficiency,%
\begin{equation}
IQE=\frac{\beta N_{2d}^{2}}{J/e}=1-2\frac{J_{0}}{J}\left[ \sqrt{\frac{J}{%
J_{0}}+1}-1\right]  \label{19}
\end{equation}%
where $J_{0}=e\gamma _{b}^{2}\left( 4\beta \right) ^{-1}$ and $%
A/A_{b}=N_{qw}h/h_{b}$ is assumed to simplify the above expressions.

\begin{figure}[tbp]
\centering
\includegraphics[width=0.5\textwidth]{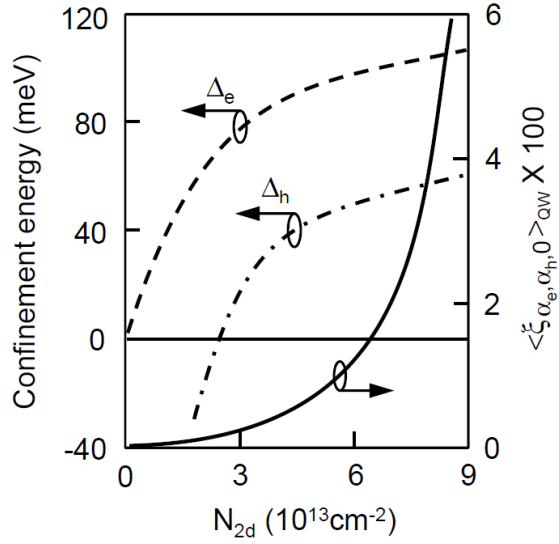}
\caption{Average QW confinement energies (left axis) and electron-hole
wavefunction overlap (right axis) versus carrier density. The curves are
extracted from solving $k\cdot p$ and Poisson equations. A negative average
hole confinement energy is possible because of the tilt in QW confinement
potential and the presence of states in the outer barrier regions cladding
the QWs, as shown in Fig.~\protect\ref{fig:4}(a).}
\label{fig:2}
\end{figure}

Bandstructure input to Eq. (\ref{19}) are the confinement energies $\Delta
_{e}$, $\Delta _{h}$ and the QW $B$ coefficient as functions of total
carrier density, $N_{2d}$. The information is extracted from the same
bandstructure calculations performed for the more comprehensive $k-$resolved
model, with the exception that only the zone center ($k_{\bot }=k=0$) values
are used. Confinement energies are approximated by $\Delta _{\sigma
}=\left\langle \varepsilon _{\sigma ,\alpha _{\sigma },0}\right\rangle
_{QW}-\varepsilon _{\sigma ,0}^{b}$, where $\left\langle {}\right\rangle
_{QW}\ $indicates an average over QW states. Based on Eqs. (\ref{10a}), (\ref%
{11a}) and (\ref{13}), the assumption $B=\left\langle \xi _{\alpha
_{e},\alpha _{h},0}\right\rangle _{QW}\eta B_{b}$ is made, where $%
\left\langle \xi _{\alpha _{e},\alpha _{h},0}\right\rangle _{QW}$ is the
average envelop function overlap of the allowed QW transitions and $\eta $
is introduced to account for the difference in QW and barrier densities of
states. This difference is automatically taken care of in the $k-$resolved
model based on Eqs. (\ref{10}) and (\ref{11}). $\Delta _{e}$, $\Delta _{h}$
and $\left\langle \xi _{\alpha _{e},\alpha _{h},0}\right\rangle _{QW}$
versus carrier density $N_{2d}$ are plotted in Fig.~\ref{fig:2}. The sheet
(2-d) density $N_{2d}$ is for a heterostructure consisting 5 QWs and 6
barrier layers that totals 84nm in width.

\begin{figure}[tbp]
\centering
\includegraphics[width=0.5\textwidth]{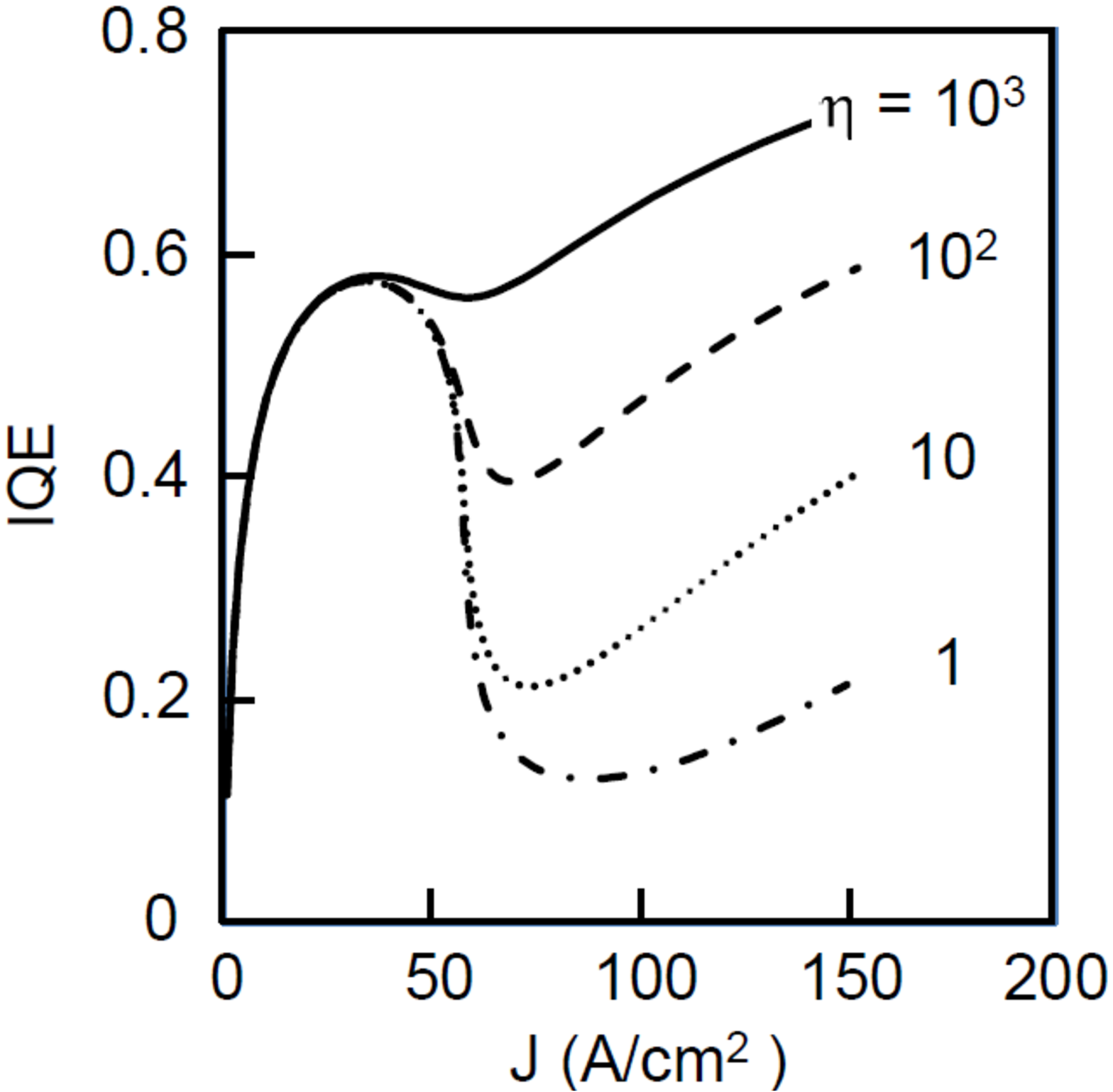}
\caption{Internal quantum efficiency versus current density computed using
Eq. (\protect\ref{19}) from an extended $ABC$ model. The model isolates the
bandstructure-induced droop mechanism. The curves are for different $\protect%
\eta $, a free parameter accounting for difference between QW and barrier
bimolecular radiative recombination coefficients ($B$ and $B_{b}$,
respectively) because of differences in densities of state.}
\label{fig:3}
\end{figure}

Figure~\ref{fig:3} shows IQE versus current density computed with Eq. (\ref%
{19}) and for different $\eta $. Input parameters are $T=300K$, $%
N_{qw}h/h_{b}=0.16$ and $A_{b}^{2}/B_{b}=1.1\times 10^{24}cm^{-3}s^{-1}$.
All the curves depict pronounced efficiency droop from the extended $ABC$
model, where carrier dependences of confinement energies and QW bimolecular
radiative coefficient are taken into account. They also indicate that the
appearance of droop is insensitive to the fitting parameter $\eta $, which
affects only the IQE recovery arising from increase in QW emission.

While the above exercise reveals that bandstructure changes is the source of
droop, differences between Figs.~\ref{fig:1} and~\ref{fig:2} suggest that
there is also influence from other contributions. That experimental results
are in closer agreement with Fig.~\ref{fig:1} indicates the importance of
these contributions in present LEDs. They include energy dispersions,
carrier leakage and nonequilibrium carrier effects, such as an incomplete
transfer of the carrier population from barrier to QW states because of
finite intraband collision rates. The presence of nonequilibrium effects is
verified from least-squares fits of computed carrier populations to
Fermi-Dirac distributions. For $J=150A/cm^{2}$, the fits indicate elevated
plasma temperatures of $T>360K$ for carrier-phonon scattering rate $\gamma
_{c-p}=10^{13}s^{-1}$ and $T>600K$ for $\gamma _{c-p}=10^{12}s^{-1}$.

Lastly, the dynamical solution gives the carrier densities in QW and barrier
states. The conversion to bulk (3-d) density is via division by the total QW
layer width $N_{qw}h$ in the case of the QW and by the total barrier width $%
h_{b}$ in the case of the barrier. When performing the bandstructure
calculation, quasiequilibrium condition is assumed to determine the QW and
barrier bulk densities used in the solution of Poisson equation. This is an
inconsistency that is acceptable provided the dynamical solution does not
produce carrier distributions deviating too far from quasiequilibrium
distributions. Even though the current density versus carrier density
relationship depends on the input to the dynamical problem, and therefore,
different for the different curves in Fig. 1, some insight into the
connection between bandstructure and IQE excitation dependence may be
obtained by examining Figs. 1 and 2 together. The onset of droop in the
curves in Fig. 1 occurs around $45A/cm^{2}$, which corresponds to $N_{2d}$
around $10^{13}cm^{-2}$ or a 3-d QW carrier density of $1\times $ to $%
1.2\times 10^{18}cm^{-3}$. At these densities, the QCSE is essentially
unscreened. At the start of IQE recovery which occurs over the range of $60$
to $120A/cm^{2}$, the corresponding carrier densities are $2.8\times
10^{13}<N_{2d}<3.0\times 10^{13}cm^{-2}$ or 3-d QW carrier density of $%
8.5\times $ to $9\times 10^{18}cm^{-2}$. According to Fig. 2, these are
densities where wavefunction overlap is no longer negligible. Between the
IQE peak and recovery, $N_{2d}$ changes from approximately $10^{13}$ to $%
3.0\times 10^{13}cm^{-2}$. Within that carrier density range, Fig. 2 shows
significant increase in QW-barrier electron and hole energy separations.

\section{Discussion of results}

\begin{figure}[tbp]
\centering
\includegraphics[width=0.5\textwidth]{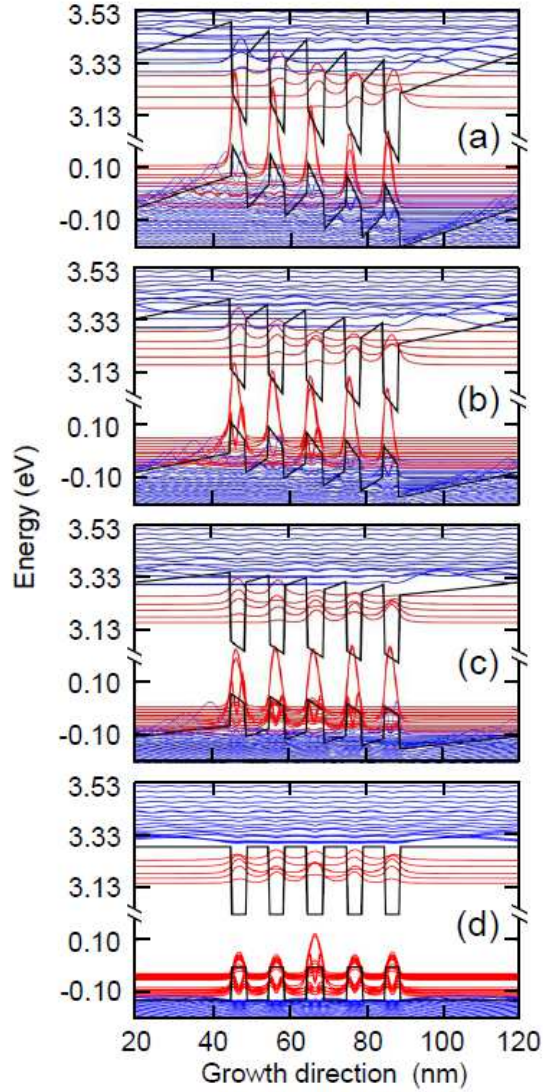}
\caption{Absolute square of envelop functions for electrons and holes for
carrier densities, $N_{2d}=$ (a) $2.25\times ,$ (b) $3.47\times $ and (c) $%
6.89\times 10^{13}cm^{-2}$. Figure~\protect\ref{fig:4}(d) is the flat-band
limit. Each curve is displaced according to its bandedge energy for clarity.
Envelop functions belonging to QW and barrier states are indicated by red
and blue curves, respectively. The black lines plot the confinement
potentials. The x-axis is along the growth direction.}
\label{fig:4}
\end{figure}

Further insight into the bandstructure-induced droop mechanism is possible
from closer examination of the bandstructure changes with excitation. Figure~%
\ref{fig:4} shows the absolute square of electron and hole envelop functions
at zone center ($k_{\bot }=k=0$) for four different carrier densities. For
clarity, the curves are separated vertically according to their associated
energies. The black lines plot the electron and hole confinement potentials,
while the red and blue curves indicate the QW and barrier states,
respectively.

Starting at a carrier density of $N_{2d}=2.3\times 10^{13}cm^{-2}$, Fig.~\ref%
{fig:4}(a) depicts confinement potentials differing appreciably from the
flat-band situation [see Fig.~\ref{fig:4}(d)]. A result is small energy
separation between QW and barrier states, leading to comparable QW and
barrier populations, especially for the holes. Optical emission from barrier
transitions occur via the contribution $\sum_{k}b_{k}n_{e,k}^{b}n_{h,k}^{b}$%
, as soon as the product of electron and hole populations, $%
n_{e,k}^{b}n_{h,k}^{b}$ becomes nonzero. In contrast, the QW contribution $%
\sum_{\alpha _{e},\alpha _{h},k_{\perp }}b_{\alpha _{e},\alpha _{h},k_{\bot
}}n_{e,\alpha _{e},k_{\perp }}n_{h,\alpha _{h},k_{\perp }}$ is negligible,
even though the product $n_{e,\alpha _{e},k\bot }n_{h,\alpha _{h},k\bot }$
may be appreciable. This is because QCSE spatially separates electrons and
holes in the QWs, resulting in very small dipole matrix elements for QW
transitions.

At a higher carrier density of $N_{2d}=3.4\times 10^{13}cm^{-2}$, increased
screening of QCSE leads to higher energy separation between QW and barrier
states as shown in Fig.~\ref{fig:4}(b). This causes the barrier populations
to decrease relative to those of the QW. However, the QCSE is still
sufficient to suppress the dipole matrix element. The net result is reduced
IQE because the smaller increase in $\sum_{k}b_{k}n_{e,k}^{b}n_{h,k}^{b}$
with increasing excitation that is not compensated by a corresponding
increase in $\sum_{\alpha _{e},\alpha _{h},k_{\perp }}b_{\alpha _{e},\alpha
_{h},k_{\bot }}n_{e,\alpha _{e},k_{\perp }}n_{h,\alpha _{h},k_{\perp }}$.
Important to the appearance of droop is a lag between the increase in
confinement energies and the increase in QW dipole matrix element, as
illustrated in Fig.~\ref{fig:2} within the region $2.5\times
10^{13}cm^{-2}<N_{2d}<5\times 10^{13}cm^{-2}$.

For the increase in QW emission, a high carrier density is necessary to
sufficiently screen the QW electric field. That is the case for Fig. ~\ref%
{fig:4}(c), where $N=6.8\times 10^{13}cm^{-2}$. An appreciable QW emission
leads to a reversal of the IQE droop as shown in Figs.~\ref{fig:1} and ~\ref%
{fig:3}. Lastly, Fig.~\ref{fig:4}(d) shows the asymptotic flat-band case,
both for reference and as a guide for assigning QW and barrier states. Note
that some ambiguity remains, especially with the $n=2$ subbands, which lie
mostly in the triangular barrier regions of the confinement potentials at
finite carrier densities.

Questions remain concerning the bandstructure-induced droop mechanism. For
example, one might expect a significant red shift of emission energy when
optical transitions changes from barrier dominated to QW dominated. This
need not be the case because of the energy level shifts associated with the
QCSE and Franz-Keldysh effects \cite{keldysh}.

\begin{figure}[tbp]
\centering
\includegraphics[width=0.5\textwidth]{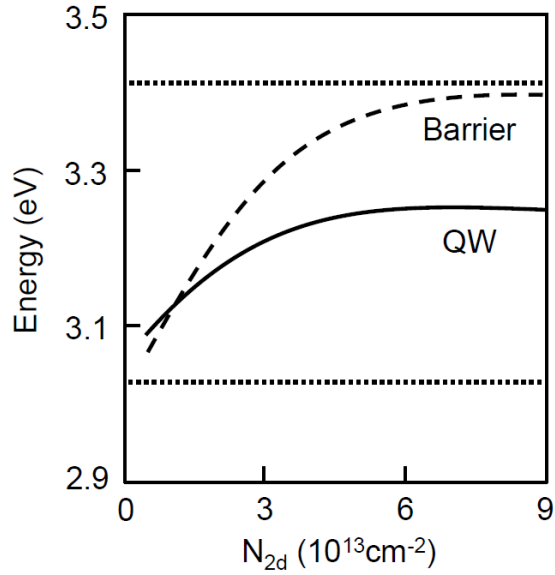}
\caption{Average QW and barrier bandedge energies (solid and dashed curves,
respectively) versus carrier density. Optical emission should be centered
approximately at the lower of the 2 curves. The upper and lower dotted lines
indicate the strained-InGaN and unstrained-GaN bulk bandgap energies.}
\label{fig:5}
\end{figure}

The curves in Fig.~\ref{fig:5} show the carrier density dependences of the
average QW and barrier bandedges, $\left\langle \varepsilon _{e,\alpha
_{e},0}\right\rangle _{QW}+\left\langle \varepsilon _{h,\alpha
_{h},0}\right\rangle _{QW}$ and $\varepsilon _{e,0}^{b}+\varepsilon
_{h,0}^{b}$, respectively. To a good approximation, emission energy is
centered around the lower of the curves, which means that except for slight
deviations around the cross-over region, the emission energy is blue shifted
with increasing excitation. Furthermore, it is always below the zero-field
barrier bandgap.

Another question concerns the curves depicting IQE recovery at current
densities lower than observed in present experiments. This discrepancy
suggests the presence of other loss mechanisms, such as Auger carrier loss.

\begin{figure}[tbp]
\centering
\includegraphics[width=0.5\textwidth]{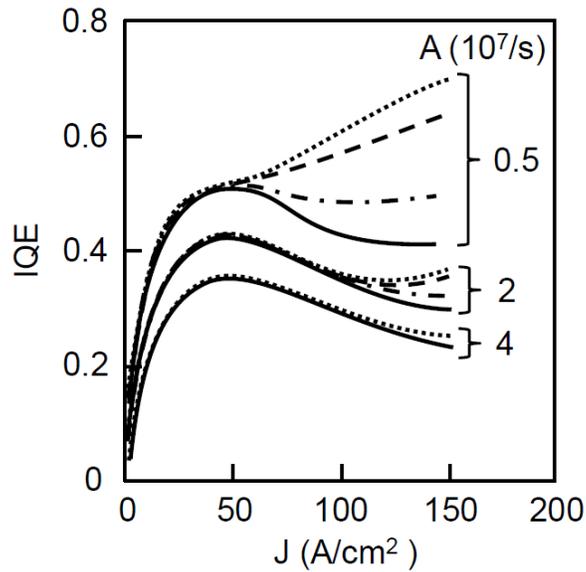}
\caption{Internal quantum efficiency versus current density showing the
influence of Auger carrier loss for different QW SRH coefficients. The Auger
coefficients are $C=0$, $10^{-32}$, $5\times 10^{-32}$ and $%
10^{-31}cm^{6}s^{-1}$ (dotted, dashed, dot-dashed and solid curves,
respectively).}
\label{fig:6}
\end{figure}

To illustrate the effect of Auger scattering, Auger carrier loss is
incorporated into Eqs. (\ref{10}) and (\ref{11}), as described in Ref. \cite%
{chow1}, and the results are shown in Fig.~\ref{fig:6} for $A/A_{b}=0.5,$ $2$
and $4$, with Auger coefficient $C=0,$ $10^{-32}$, $5\times 10^{-32}$ and $%
10^{-31}cm^{6}s^{-1}$ (dotted, dashed, dot-dashed and solid curves,
respectively). For clarity, the $A/A_{b}=1$ case in Fig.~\ref{fig:1} is
omitted. The curves shows the prolonging of the efficiency droop by Auger
carrier loss. More importantly, the necessary Auger coefficient is shown to
be $C\gtrsim 5\times 10^{-32}cm^{6}s^{-1}$ which is appreciably smaller than
that used in $ABC$ models and are within the range predicted by microscopic
calculation \cite{dellaney}.

\section{Summary}

This paper describes an approach to modeling InGaN LEDs that involves the
self-consistent solution of bandstructure and carrier population problems.
The motivation is to provide direct input of bandstructure properties, in
particular, their carrier-density dependences arising from screening of
piezoelectric and spontaneous polarization fields. Other advantages include
consistent treatment of spontaneous emission, carrier capture and leakage
and nonequilibrium effects, as well as description of optical emission from
quantum-well and barrier transitions on equal footing.

The approach is applied to investigate the internal quantum efficiency as a
function of injection current for a multi-QW InGaN LED. Among the behaviors
that result from taking into account the excitation dependences of
bandstructure, is a possible contribution to efficiency droop. Simulations
performed with different sets of input parameter values show robustness of
this mechanism.\ A simple, back-of-the-envelop derivation is used to trace
the droop mechanism to bandstructure changes from screening of the
quantum-confined Stark effect. Basically, the initial IQE peak has emission
contribution from barrier states. The droop is caused by carrier transfer
from these barrier states to QW states, where emission strength is weaker
because the quantum-confined Stark effect remains largely unscreened.

Lastly, the results presented in this paper should not be generalized to
imply that bandstructure effects are responsible for the entire droop
phenomenon in all experiments. In fact, simulation results are presented to
demonstrate the importance of other contributions, both intrinsic and
extrinsic, for describing present IQE experiments. It is possible that the
differences in observed droop behavior (involving different LED emitting
wavelengths, polar versus non polar substrates, with or without electron
blocking layers, etc.) arise from differences in the relative importance of
various mechanisms. The $k$-resolved LED model described in this paper can
provide a more accurate estimation of their relative strengths. Furthermore,
as shown in Sec. 4 for Auger loss, the model can also put arguments drawn
from experimental-curve fitting on firmer ground by providing better
connection to first-principles theory.

\textbf{Acknowledgement\smallskip }

This work is performed at Sandia's Solid-State Lighting Science Center, an
Energy Frontier Research Center (EFRC) funded by the U. S. Department of
Energy, Office of Science, Office of Basic Energy Sciences. The author
thanks A. Armstrong, M. Crawford, P. Smowton and J. Tsao for helpful
discussions.



\begin{thebibliography}{99}
\bibitem{krames} M.R. Krames, O.B. Shchekin, R. Mueller-Mach, G.O. Mueller,
L. Zhou, G. Harbers and M. G. Craford, "Status and future of high-power
light-emitting diodes for solid-state lighting," J. Display Technology 
\textbf{3}, 160-175 (2007).

\bibitem{schubert} M. H. Kim, M. F. Schubert, Q. Dai, J. K. Kim, E. F.
Schubert, J. Piprek, and Y. Park, "Origin of efficiency droop in GaN-based
light-emitting diodes," Appl. Phys. Lett. \textbf{91}, 183507-183510 (2007).

\bibitem{shen} Y.C. Shen, G.O. M\"{u}ller, S. Watanabe, N.F. Gardner, A.
Munkholm and M.R. Krames, "Auger recombination in InGaN measured by
photoluminescence," Appl. Phys. Lett. \textbf{91}, 141101-141101 (2007).

\bibitem{efremov} A. A. Efremov, N. I. Bochkareva, R. I. Gorbunov, D. A.
Larinvovich, Yu. T. Rebane, D. V. Tarkhin and Yu. G. Shreter, "Effect of the
joule heating on the quantum efficiency and choice of thermal conditions for
high-power blue InGaN/GaN LEDs," Semiconductors \textbf{40}, 605 (2006).

\bibitem{chichibu} S. F. Chichibu, T. Azuhata, M. Sugiyama, T. Kitamura, Y.
Ishida, H. Okumurac, H. Nakanishi, T. Sota and T. Mukai, "Optical and
structural studies in InGaN quantum well structure laser diodes," J. Vac,
Sci. Technol. B \textbf{19}, 2177 (2001).

\bibitem{hader1} J. Hader, J.V. Moloney and S.W. Koch, "Density-activated
defect recombination as a possible explanation for the efficiency droop in
GaN-based diodes," Appl. Phys. Lett. \textbf{96}, 221106-221108 (2010).

\bibitem{ryu} H.-Y Ryu, H.-S. Kim and J.-I. Shim, "Rate equation analysis of
efficiency droop in InGaN light-emitting diodes," Appl. Phys. Lett. \textbf{%
95}, 081114-081117 (2009).

\bibitem{hader2} J. Hader, J.V. Moloney, B. Pasenow, S.W. Koch, M. Sabathil,
N. Linder and S. Lutgen, "On the important of radiative and Auger losses in
GaN-based quantum wells," Appl. Phys. Lett. \textbf{92}, 261103-261105
(2008).

\bibitem{dellaney} K.T. Dellaney, P. Rinke and C.G. Van de Walle, "Auger
recombination rates in nitrides from first principles," Appl. Phys. Lett. 
\textbf{94}, 191109-191111 (2009).

\bibitem{shur} A. Bykhovshi, B. Gelmonst and M. Shur, "The influence of the
strain-induced electric field on the charge distribution in GaN-AlN-GaN
structure," J. Appl. Phys. \textbf{74}, 6734-6739 (1993).

\bibitem{hangleiter} J. S. Im, H. Kollmer, J. Off, A. Sohmer, F. Scholz and
A. Hangleiter, "Reduction of oscillator strength due to piezoelectric fields
in GaN/AlGaN quantum wells," Phys. Rev. B \textbf{57}, R9435-R9438 (1998).

\bibitem{chow1} W. W. Chow, M. H. Crawford, J. Y. Tsao and M. Kneissl,
"Internal efficiency of InGaN light-emitting diodes: Beyond a
quasiequilibrium model," Appl. Phys. Lett. \textbf{97}, 121105-121107 (2010).

\bibitem{chuang} S. L. Chuang and C. S. Chang, "$k\cdot p$ method for
strained wurtzite semiconductors," Phys. Rev. B \textbf{54}, 2491-2504
(1996).

\bibitem{jaynes} E. Jaynes and F. Cummings, "Comparison of quantum and
semiclassical radiation theories with application to the beam maser*Proc.
IEEE \textbf{51}, 89-109 (1963).

\bibitem{chow2} W.W. Chow, H.C. Schneider, S.W. Koch, C.H. Chang, L.
Chrostowski and C. J. Chang-Hasnain, "Nonequilibrium model for semiconductor
laser modulation response," IEEE Journ. Quantum Electron. \textbf{38}, 402
(2002).

\bibitem{waldmueller} I. Waldmueller, W. W. Chow, M. C. Wanke and E. W.
Young, "Non-equilibrium many-body theory of intersubband lasers," IEEE
Journ. Quantum Electron. \textbf{42}, 292-301 (2006).

\bibitem{chow3} W. W. Chow, A. F. Wright, A. Girndt, F. Jahnke and S. W.
Koch, "Microscopic theory of gain for an InGaN/AlGaN quantum well laser,"
Appl. Phys. Lett. \textbf{71}, 2608-2610 (1997).

\bibitem{book} W. W. Chow and S. W. Koch, \textit{Semiconductor-Laser
Fundamentals: Physics of the Gain Materials} (Springer, Berlin, 1999).

\bibitem{jenkins} S. J. Jenkins, g. P. Srivastava and J. C. Inkson, "Simple
approach to self-energy corrections in semiconductors and insulators," Phys.
Rev. B \textbf{48}, 4388 (1993).

\bibitem{wright} A. F. Wright and J. S. Nelson, "Consistent structural
properties for AlN, GaN, and InN," Phys. Rev. B \textbf{51}, 7866-7869
(1995).

\bibitem{wei} S. H. Wei and A. Zunger, "Valence band splittings and band
offsets of AlN, GaN, and InN," Appl. Phys. Lett. \textbf{69}, 2719-2711
(1996).

\bibitem{ambacher} O. Ambacher, "Growth and applications of Group
III-nitrides," J. Phys. D: Appl. Phys. \textbf{31}, 2653-2710 (1998).

\bibitem{andy} A. Armstrong, Sandia National Laboratories, Albuquerque, NM
87185 (personal communication, 2010).

\bibitem{keldysh} L. V. Keldysh, "Behaviour of non-metallic crystals in
strong electric fields,," Soviet Physics JETP \textbf{6,} 763-770 (1958).
\end{thebibliography}
\end{document}